\documentclass[11pt,a4paper]{article}
\usepackage{jcappub}

\usepackage{bm}
\usepackage{epsfig}
\usepackage{citesort}
\usepackage{graphicx}
\usepackage{amssymb}
\usepackage{color}
\usepackage{subfigure}

\newcommand{\alt}{\mbox{\;\raisebox{.3ex}
  {$<$}$\!\!\!\!\!$\raisebox{-.9ex}{$\sim$}\;}}
\newcommand{\agt}{\mbox{\;\raisebox{.3ex}
  {$>$}$\!\!\!\!\!$\raisebox{-.9ex}{$\sim$}}\;}

\renewcommand\({\left(}
\renewcommand\){\right)}
\renewcommand\[{\left[}
\renewcommand\]{\right]}

\renewcommand\({\left(}
\renewcommand\){\right)}
\renewcommand\[{\left[}
\renewcommand\]{\right]}
\newcommand{\be}{\begin{equation}}
\newcommand{\ee}{\end{equation}}
\newcommand{\bea}{\begin{eqnarray}}
\newcommand{\eea}{\end{eqnarray}}

\newcommand{\parte}{{\ttfamily PArthENoPE}}

\begin{document}



\subheader{\hfill MPP-2010-148}

\title{
Cosmological bounds on\\ sub-MeV mass axions}

\author[a]{Davide~Cadamuro}
\author[b]{Steen~Hannestad}
\author[a]{Georg~Raffelt}
\author[a]{Javier~Redondo}

\affiliation[a]{Max-Planck-Institut f\"ur Physik
(Werner-Heisenberg-Institut)\\
F\"ohringer Ring 6, D-80805 M\"unchen, Germany}

\affiliation[b]{Department of Physics and Astronomy\\
University of Aarhus, DK-8000 Aarhus C, Denmark}

\emailAdd{cadamuro@mppmu.mpg.de}
\emailAdd{sth@phys.au.dk}
\emailAdd{raffelt@mppmu.mpg.de}
\emailAdd{redondo@mppmu.mpg.de}

\abstract{Axions with mass $m_a\agt0.7$~eV are excluded by
cosmological precision data because they provide too much hot dark
matter. While for $m_a\agt20$~eV the $a\to 2\gamma$ lifetime drops
below the age of the universe, we show that the cosmological
exclusion range can be extended to $0.7~{\rm eV}\alt
m_a\alt300$~keV, primarily by the cosmic deuterium abundance: axion
decays would strongly modify the baryon-to-photon ratio at BBN
relative to the one at CMB decoupling. Additional arguments include
neutrino dilution relative to photons by axion decays and spectral
CMB distortions. Our new cosmological constraints complement
stellar-evolution and laboratory bounds.}

\maketitle

\section{Introduction}                        \label{sec:introduction}

The era of precision cosmology has opened new opportunities to use
the universe as a particle-physics laboratory. Cosmological neutrino
mass limits and the prospect of eventually measuring this
fundamental parameter in the sky is, of course, the most notable
example~\cite{Lesgourgues:2006nd,Hannestad:2006as}. Analogous
arguments allow to derive limits on other hypothetical  low-mass
particles and here axions are perhaps the most interesting
case~\cite{Hannestad:2003ye,Hannestad:2005df,Melchiorri:2007cd,%
Hannestad:2007dd,Hannestad:2008js}, where cosmological data
imply $m_a<0.7$~eV \cite{Hannestad:2010yi}. Of course, if axions are
the cold dark matter of the universe~\cite{Sikivie:2006ni}, they are
very light ($m_a\alt10~\mu$eV) and such arguments are moot.
Therefore, the hot dark matter limits are primarily useful to
complement stellar energy-loss arguments~\cite{Raffelt:2006cw} and
the direct search for solar axions by CAST at CERN~\cite{Arik:2008mq}
and the Tokyo axion helioscope~\cite{Inoue:2008zp}.

Axions decay by $a\to 2\gamma$, the lifetime dropping below the age
of the universe for $m_a\agt20$~eV. For a sufficiently large mass,
the decay happens early enough to thermalize the secondary photons
and hot-dark matter or overclosure arguments no longer apply.

However, decaying axions are still constrained by several cosmological
arguments.
Before the cosmological precision era, such questions were discussed
in great detail for axion-like particles~\cite{Masso:1995tw,Masso:1997ru}.
A central argument was
the impact of excess radiation provided by axions during big-bang
nucleosynthesis (BBN).  Today it has become clear that BBN does not
credibly exclude excess radiation provided by one additional thermal
degree of freedom and some amount of excess radiation may even be
favoured~\cite{Aver:2010wq,Izotov:2010ca}. Moreover, cosmological
precision data somewhat prefer excess radiation equivalent to 1--2
neutrino species at the epoch of CMB
decoupling~\cite{Hamann:2007pi,Hamann:2010pw,Komatsu:2010fb,GonzalezGarcia:2010un,Hamann:2010bk}.
Therefore, we here address the question which axion mass range
beyond 0.7~eV is actually excluded by cosmological arguments alone.

If axions decay not too early, spectral
distortions of the CMB provide restrictive constraints~\cite{Masso:1997ru}.
For earlier decays (larger masses), the decay photons thermalize and heat the
CMB relative to cosmic neutrinos. Since all cosmological quantities
are normalized to the {\em measured\/} CMB temperature, this means
that effectively neutrinos are diluted. If axions decay
non-relativistically and thus out of equilibrium, this effect can be
large and leads to a lowered radiation density at CMB decoupling,
increasing the tension with precision data that actually favour a
radiation excess. Most importantly, the baryon density at CMB
decoupling is also diluted.
Therefore, the baryon abundance relevant for BBN is larger than
implied by CMB data, leading to a smaller primordial deuterium
abundance. Confronting this with observations requires that axions
are not significantly abundant during BBN.
This new BBN limit requires $m_a\agt300$~keV.  
Our new bounds therefore nicely overlap with the constraints from beam dump and 
reactor experiments~\cite{Bergsma:1985qz,Konaka:1986cb,Riordan:1987aw,Bross:1989mp,Altmann:1995bw,Chang:2006ug}.

To derive this result we first describe the relevant phenomenological properties of axions in 
Sec.~\ref{sec:axions}.  In Sec.~\ref{sec:freezeout}, we discuss the
modification of the cosmic expansion history and the relative
contributions of different forms of radiation in our decaying axion
scenario. Constraints based on BBN, neutrino dilution, and CMB
distortions are presented in
Secs.~\ref{sec:baryons}--\ref{sec:distortions} before concluding in
Sec.~\ref{sec:conclusions}.

\section{Axion properties}                          \label{sec:axions}

The axion arises in the Peccei-Quinn solution of the strong CP
problem as the Nambu-Goldstone boson of a new global spontaneously
broken U(1)$_{\rm PQ}$ axial symmetry~\cite{Peccei:2006as}. The main
ingredient of the PQ mechanism is that U(1)$_{\rm PQ}$ is colour
anomalous, providing for an axion-gluon interaction of the form
\begin{equation}
\mathcal{L}_{a G G} =\frac{a}{f_a}\frac{\alpha_s}{8\pi}\,
G_a^{\mu\nu}\tilde{G}_{a\mu\nu}\,.
\end{equation}
The energy scale $f_a$, the axion decay constant, is the main
parameter that fixes low-energy axion phenomenology. This term
automatically cancels the CP-violating $\theta$ term of QCD because
it generates nonperturbatively a potential for the axion, driving
the axion field to the CP-conserving position. The axion mass thus
created is
\begin{equation}
m_a=\frac{f_\pi m_\pi}{f_a}\frac{\sqrt{z}}{1+z}\simeq 6~{\rm eV}
\(\frac{10^6~\rm GeV}{f_a}\),
\end{equation}
where $m_\pi=135$~MeV is the neutral pion mass, $f_\pi=92$~MeV its
decay constant, and $z=m_u/m_d$. We use the canonical value $z=0.56$,
although the possible range is $z=0.35$--0.6~\cite{Nakamura:2010zzi},
implying a 10\% uncertainty of the axion mass.

In the minimal scenario we shall consider, axions do not have tree-level interactions with
leptons. There are thus called \emph{hadronic} axions. 
The freeze-out of sub-eV hadronic axions is determined by
their interaction with pions. However, we here consider much larger
masses and much later freeze-out. Their only relevant interaction is
with photons,
\begin{equation}\label{lagrangian}
\mathcal{L}_{a\gamma\gamma}=
\frac{g_{a\gamma}}{4}F^{\mu\nu}\tilde{F}_{\mu\nu}a
=-g_{a\gamma}{\bf E}\cdot{\bf B}\,a\,,
\end{equation}
where $F$ is the electromagnetic field-strength tensor, $\tilde F$
its dual, and ${\bf E}$ and ${\bf B}$ the electric and magnetic
field, respectively. The coupling constant
\begin{equation}
g_{a\gamma}=
\frac{\alpha}{2\pi f_a}\left(\frac{E}{N}-\frac{2}{3}\frac{4+z}{1+z}\right)
\equiv \frac{\alpha}{2\pi f_a}\,1.9\ \delta\,,
\end{equation}
consists of a model-independent contribution from $a$-$\pi$-$\eta$
mixing and a model-dependent one parameterized by the ratio $E/N$ of
the U(1)$_{\rm PQ}$ electromagnetic and colour anomalies. The uncertainties
from the up/down quark-mass ratio $z$ and from the model-dependence
of $E/N$ can be lumped into the parameter $\delta$.
We have normalized $\delta$ such that $\delta=1$ for $E=0$, happening for instance
in the KSVZ model~\cite{Kim:1979if, Shifman:1979if}.

The well-known $a\to2\gamma$ decay
rate is
\begin{equation}
\label{decayrate}
\Gamma_{a\to\gamma\gamma}= \frac{g_{a\gamma}^2 m_a^3}{64 \pi}\simeq
1.1\times10^{-24}~\mbox{s}^{-1}\,
{\(\frac{m_a}{\mbox{eV}}\)}^5\delta^2.
\end{equation}
For $\delta=1$, axions with $m_a<18$~eV live longer than the age of
the universe. In this paper, we shall consider $\delta$ values of order one,
excluding radical fine-tuning or very large $E/N$ scenarios.
Note however, than a value $\delta=0$ provides the most
restrictive limit: axions would not disappear and eventually over-close the
universe.

In extended models such as the DFSZ~\cite{Zhitnitsky:1980tq,Dine:1981rt}, axions also have tree-level interactions with leptons. 
The Nambu-Goldstone boson nature of axions determines the form of the interaction to be 
of the derivative axial-current form~\cite{Gelmini:1982zz} 
\begin{equation}
\label{axionlepton}
{\cal L}_l = \frac{C_l}{2 f_a} \, \bar \psi_l \gamma^\mu \gamma_5 \psi_l\, (\partial_\mu a)
\equiv \frac{C_l m_l}{f_a} \, \bar \psi_l \gamma_5 \psi_l\, a  , 
\end{equation}
where the $C_l$'s are model-dependent numerical coefficients . 
Note that the coupling is proportional to the lepton mass\footnote{The equivalence between the axial and pseudoscalar forms of Eq.~\eqref{axionlepton} is not always guaranteed but it will be so for the applications of this paper where we consider processes involving the emission or absorption of one Nambu-Goldstone boson at a time.}.
For late cosmology we will be mostly interested in the coupling to electrons since the coupling 
to neutrinos is extremely suppressed by the small neutrino masses.

\section{Decaying axion cosmology}               \label{sec:freezeout}


Our main cosmological arguments depend on modified baryon-to-photon
and neutrino-to-photon ratios at various epochs. One difficulty is
that the freeze-out epoch of axions in the relevant mass range can
coincide with the $e^+e^-$ annihilation epoch and with the BBN epoch
itself. Moreover, for $m_a\alt20$~keV, axions first decouple from the
Primakoff process and then recouple by inverse decays, causing a
somewhat complicated evolution of the radiation fields. As a first
step we therefore discuss this evolution.

Thermal axions are produced after the QCD epoch by hadronic
processes such as $\pi + \pi \leftrightarrow \pi +a$. As the
universe cools, hadrons disappear and the photo-production of axions from electrons, $e^\pm\gamma\to e^\pm a$, dominate.
For hadronic axions, the most relevant reaction is the Primakoff process (see Fig.~\ref{fig8a} in Appendix~\ref{app:freezeout}). 
A simple estimate for $T\gg m_a$ is
\begin{equation}
\Gamma_P \sim \alpha g_{a\gamma}^2 n_e  , 
\end{equation}
where $n_e$ is the density of electrons plus positrons. 
For non purely hadronic axions, the direct axion-electron coupling allows for the Compton process 
(see Fig.~\ref{fig8b} in Appendix~\ref{app:freezeout}).
An estimate of its rate is
\begin{equation}
\Gamma_C \sim \alpha \frac{g_{ae}^2}{{\rm max}\{T^2,m_e^2\}} n_e . 
\end{equation}
At temperatures close to the electron mass, this process is more efficient than the Primakoff since involves less powers of $\alpha$, and when present will dominate the axion thermalization.
Other processes like $e^+ e^-\to \gamma a$ are subdominant and can be neglected in the following. 
More details about the Primakoff and Compton processes are given in Appendix~\ref{app:freezeout}.

\begin{figure}[tbp] 
   \centering
   \includegraphics[width=8cm]{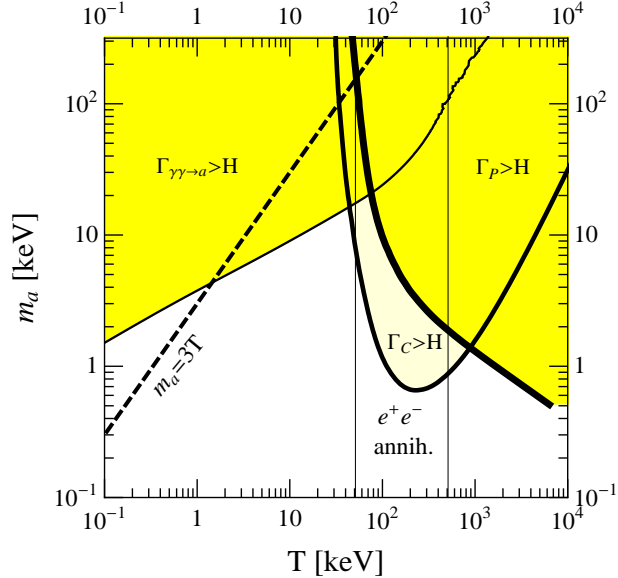}
   \caption{Axion decoupling and recoupling ($\delta=1$, $C_e=1/6$).
   Thick solid line: Freeze-out of Primakoff process. 
   Medium solid line: Coupling and freeze-out of the Compton process.
   Thin solid line: Recoupling of inverse decay. In the yellow shaded region,
   axions are in thermal equilibrium, where the lighter yellow region is only relevant if the Compton process is effective. The dashed diagonal line
   denotes $m_a=3T$.
    The vertical lines delimit the
   $e^+e^-$ annihilation epoch.}
   \label{fig1}
\end{figure}

The Primakoff and/or Compton processes freezes out when their rate drops below the cosmic expansion rate
\begin{equation}\label{eq:Hubble}
H = \frac{1}{R}\frac{dR}{dt}=
\sqrt{\frac{8 \pi}{3
m_{\rm Pl}^2} \rho }\equiv
\sqrt{\frac{8 \pi}{3
m_{\rm Pl}^2} \frac{\pi^2}{30}g_*}\;T^2\,,
\end{equation}
where $R$ is the cosmic scale factor, $m_{\rm Pl}=1.22 \times
10^{19}$~GeV is the Planck mass, $\rho$ the total energy density of the universe,
$T$ the photon temperature and
the last equality defines $g_*$, the effective number of thermally excited degrees of
freedom. The rates $\Gamma_P$ and $\Gamma_C$ decrease 
exponentially when $e^+e^-$ annihilation sets in, so the Primakoff and Compton processes
indeed freeze out (see Fig.~\ref{fig1}).

Another processes based on the two-photon vertex are decay and inverse
decay. For $T\gg m_a$ the inverse decay rate is
\begin{equation}
\Gamma_{\gamma\gamma\to a} \simeq
\Gamma_{a\to\gamma\gamma}\frac{m_a^2-4 m_\gamma^2}{m_a^2}
\left\langle\frac{m_a}{\omega}\right\rangle\,,
\end{equation}
where $\left\langle {m_a/\omega}\right\rangle$ is a thermally
averaged time dilatation factor. Decay or inverse decay is only
possible if $m_a>2m_{\gamma}$, where $m_\gamma$ is the plasmon mass.
For $T\gg m_e$, we have $m_\gamma=\sqrt{3\alpha \pi/2}\ T\lesssim T$ while for $T\ll m_e$ we 
have $m_\gamma \ll T$ so the decay/inverse decay channels open up not far from $m_a\sim {\rm max}\{T,m_e\}$. 
Leaving aside the threshold, for relativistic axions $\Gamma_{\gamma\gamma\to a}\propto T^{-1}$ and
this process actually recouples at a certain temperature (thin solid line in Fig.~\ref{fig1}).

In Fig.~\ref{fig1} we show the decoupling and recoupling
temperatures (horizontal axis) as a function of axion mass (vertical
axis). For a given $m_a$ the cooling universe moves on a horizontal
line from right to left, beginning with axions in thermal
equilibrium (yellow shaded). For $m_a\agt20$~keV the horizontal line
never leaves yellow territory and axions remain in thermal
equilibrium forever, first by the Primakoff process and later by
inverse decays. 
If axions couple to electrons, this region is extended until $m_a\sim 10$ keV
through the Compton process.
In this case, the cosmic axion history is governed
by thermal equilibrium. The populations of all forms of radiation
are determined by entropy conservation (adiabatic regime).

Another relatively simple range is $m_a\alt1$~keV where Primakoff
freezes out when the $e^+e^-$ plasma is fully populated and Compton
is not very effective. The
approximate temperature range of $e^+e^-$ annihilation is delimited by the two
thin vertical lines in Fig.~\ref{fig1}. In this mass range
recoupling occurs for $T\ll m_a$ as we can see in
Fig.~\ref{fig1} where the diagonal dashed line denotes
$T=3m_a$. Notice that here recoupling means axion decay
since the putative thermal axion population is exponentially Boltzmann
suppressed. Entropy is generally produced in the recoupling process, 
which in this case can be quite sizeable.
In the approximate $m_a$ range 1--5~keV, axions decouple
during the $e^+e^-$ annihilation epoch, but once more recouple in
the Boltzmann suppressed regime. For $m_a=5$--20~keV
axions recouple under relativistic conditions.

For axion-photon interaction strength $\delta\not=1$ the overall
behaviour is similar. Both the Primakoff and decay rates are
proportional to $\delta^2$ and are increasing functions of $m_a$.
Therefore, a value $\delta<1$ can be compensated by a increase in
$m_a$ and vice versa, so the decoupling and recoupling curves move
upwards when $\delta<1$ and downwards for $\delta>1$.
The same holds for the coupling to electrons $C_e$ which we have taken 
to be $1/6$ in Fig.~\ref{fig1}. 

\subsection{Primakoff decoupling and axion abundance}

\begin{figure}[t]
   \centering
   \includegraphics[width=8cm]{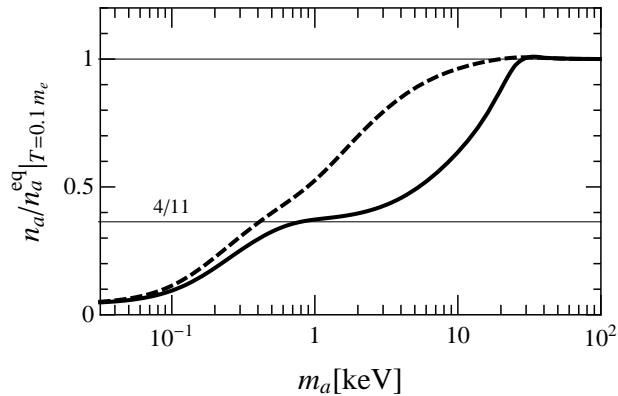}
   \caption{Axion number density $n_a$ after
   $e^+e^-$ annihilation from
   numerically solving the Boltzmann equation
   until $T=m_e/10$. The equilibrium density
   $n_a^{\rm eq}$ is defined in terms of the photon temperature.
   Solid line: only Primakoff process (hadronic axions). Dashed line: Primakoff and Compton.}
   \label{fig2}
\end{figure}

Next we must find the axion population existing after Primakoff/Compton
decoupling in those cases when axions actually leave thermal
equilibrium, i.e.\ when the cosmic evolution in
Fig.~\ref{fig1} cuts through white space. We must solve the
Boltzmann collision equation for the evolution of the axion number
density $n_a$ during the $e^+e^-$ annihilation epoch. We cast the
Boltzmann equation in the form
\begin{equation}
\label{BOLna}
\frac{d Y_a}{d\log T}=\frac{\Gamma_e+\Gamma_{\gamma\gamma\to a}}{H}\frac{d \log s}{d \log T^3} \(Y_a-Y_a^{\rm eq}\) \; , 
\end{equation}
where $Y_a=n_a/s$ and $s$ is the entropy density.
We have numerically solved for the evolution of $Y_a$ and show the
final axion abundance after $e^+e^-$ annihilation in
Fig.~\ref{fig2}. Hadronic axions with $m_a\gtrsim20$~keV indeed keep thermal
abundance. For smaller $m_a$, the final abundance is smaller because
axions decouple before the $e^+e^-$ entropy is fully released: they
end up colder than photons (which define the equilibrium
temperature). If axions couple to electrons the situation is very much the same
except that the Compton process being more effective the abundances are generally 
larger than in the purely hadronic case.

Axions modify the photon abundance in crucial ways whereas neutrinos
are unaffected. Therefore, it proves convenient to normalize
abundances in terms of the neutrino temperature which redshifts as
$T_\nu\propto R^{-1}$ after the neutrino
decoupling. So we parameterize the axion and photon abundances in terms of
their temperatures as
\begin{equation}\label{eq:AB}
A=\(\frac{T_a}{T_\nu}\)^3
\quad\hbox{and}\quad
B=\(\frac{T}{T_\nu}\)^3\,.
\end{equation}
As initial condition we use
$T_{0}=T_{\nu,0}=T_{a,0}=2$~MeV when neutrinos have
decoupled and $e^+e^-$ annihilation has not yet begun. The final
epoch is defined by $T_1=m_e/10$. Entropy conservation
during $e^+e^-$ annihilation implies
\begin{equation}
\frac{7}{2}+2+A_0=2B_1+A_1\,,
\end{equation}
where $7/2$ and 2 are the $e^+e^-$ and $\gamma$ entropy degrees of
freedom. From Fig.~\ref{fig1} we see that all of this happens
far below the dashed line, i.e.\ axions are relativistic.
We assume kinetic equilibrium even if axions decouple during the $e^+e^-$
annihilation.

If axions decouple before $e^+e^-$ annihilation like neutrinos we
have $A_1=A_0$. They do not receive any $e^+e^-$ entropy and photons
are heated by the standard amount, i.e. $B_1=\frac{11}{4}$.

On the other hand, if axions decouple during or after $e^+e^-$ annihilation they are
somewhat heated. We always have $A_0=1$ and thus
\begin{equation}
\label{B1}
B_1= \frac{13}{2(h_1+2)}
\quad\hbox{where}\quad
h=\frac{A}{B}=\(\frac{T_a}{T}\)^3
=\frac{n_a}{n_a^{\rm eq}}= \frac{s_a}{s_a^{\rm eq}}\,.
\end{equation}
It is $h$ that was plotted in Fig.~\ref{fig2}. If axions are fully
coupled during $e^+e^-$ annihilation, then $h_1=1$ and
$B_1=\frac{13}{6}$. In general we will have
$\frac{11}{4}<B_1<\frac{13}{6}$.

\subsection{Entropy production by axion decay}

Eventually axions disappear and transfer their entropy to the photon
bath, increasing the photon density both relative to neutrinos and
baryons. If axions never leave thermal equilibrium (for $m_a\agt
20$~keV), we easily find that the final photon abundance is
$\frac{13}{11}$ times the standard value. In general the entropy
transfer to photons depends on two parameters, the initial axion
abundance, parameterized by $h$, and the effectiveness of inverse
decay when axions become non-relativistic. To determine the final
photon abundance we have solved numerically for the axion phase-space
evolution (Appendix~\hbox{\ref{app:freezeout}}). Here we only
highlight the extreme cases of adiabatic decay (happening when inverse decay is
effective at the time axions become non-relativistic) and out-of-equilibrium
decay (when the inverse decay is ineffective).

If axions recouple relativistically, their temperature must catch up
with photons and the latter are cooled. In this process, radiation is
simply shuffled from one form to another and co-moving energy remains
conserved. After axion recoupling, the photon abundance is reduced to
$B_*=B_1\,[(2+h_1^{4/3})/(2+1)]^{3/4}$. Comoving entropy increases by
the factor $[(2+1)/(2+h_1)]B_*/B_1$, but this is never a big
effect. For $h_1=4/11$ the entropy increases by some 2.6\%.  Later when
axions get Boltzmann suppressed, this adiabatic process transfers
their entropy to the photon bath, heating it according to
$B_2/B_*=(2+1)/2=3/2$. Therefore, the final photon heating by axion
recoupling and adiabatic decay is given by
\begin{equation}
\frac{B_2}{B_1}=\frac{3}{2}\(\frac{2+h_1^{4/3}}{3}\)^{3/4}\,.
\end{equation}
This ratio is the number density of photons relevant at CMB
decoupling relative to the density they would have without axion
effects, where in both cases the density is measured relative to
neutrinos. Without axions $B_2/B_1=1$.

If axions recouple non-relativistically, their out-of-equilibrium
decay can produce a large amount of entropy. This point is
illustrated by an analytic approximation to the entropy
generation~\cite{Kolb:1990vq}
\begin{equation}
\frac{B_2}{B_1}=1.83\,\langle g_{*S}^{1/3}\rangle^{3/4}
\frac{m_aY_{a}(T_1)}{\sqrt{m_{\rm Pl}\Gamma_{a\to\gamma\gamma}}} =
2.8\times 10^3 \frac{h_1}{\delta} \(\frac{500\ \rm eV}{m_a}\)^{3/2}\,,
\end{equation}
where $g_{*S}$ is the effective number of entropy degrees of freedom
and $\langle g_{*S}^{1/3}\rangle$ denotes an average over the decay
time. In the last equality we have taken $g_{*S}=g_{*S}(T_1)=3.9$.

\begin{figure}[tbp]
   \centering
   \includegraphics[width=7.3cm]{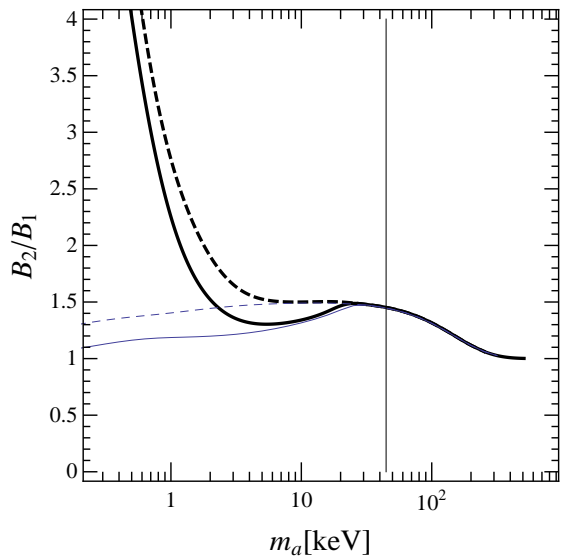}
   \caption{Photon density increase in our modified cosmology as
     expressed by $B_2/B_1$ for $\delta=1$.
     Solid and dashed lines stand for hadronic and non-hadronic 
     ($C_e=1/6$) axions respectively.
     The thin lines show the value if we assume that no
     entropy is generated in axion decay. }
   \label{fig3}
\end{figure}

For our actual BBN and neutrino dilution arguments in the following
sections we have numerically calculated the photon heating relative
to neutrinos as a function of $m_a$ by solving the Boltzmann equation
(Appendix~\ref{app:freezeout}). In Fig.~\ref{fig3} we show the
resulting $B_2/B_1$ as a function of $m_a$ 
in the case $\delta=1$.  

\section{Big-bang nucleosynthesis}                 \label{sec:baryons}

The outcome of standard BBN depends basically on the parameter $\eta=n_B/n_\gamma$, the
baryon-to-photon ratio. The overall agreement of the
predicted light-element abundances with observations implies
$5.1<\eta_{10}< 6.5$ at 95\% C.L. \cite{Nakamura:2010zzi}, where
$\eta_{10}=10^{10} \eta$. An independent determination $\eta_{10} =
6.23\pm 0.17$ derives from the CMB temperature
fluctuations~\cite{Komatsu:2010fb}. The concordance between the BBN
and CMB implied values will be disturbed in the decaying axion
cosmology because the baryon abundance is diluted after BBN. In our scenario
$\eta^{\rm BBN}/\eta^{\rm CMB}=B_2/B_1$ and therefore it follows the same curve
depicted in Fig.~(\ref{fig3}). An
increased baryon abundance relative to photons during BBN implies
than nuclear reactions are more efficient, the deuterium bottleneck
opens earlier, decreasing the residual deuterium abundance and increasing
the yield of heavier elements.
The increased expansion rate caused by the presence of axions works in the 
opposite direction but plays a sub-leading role.

\begin{figure}[tbp]
   \centering
   \includegraphics[width=15.4cm]{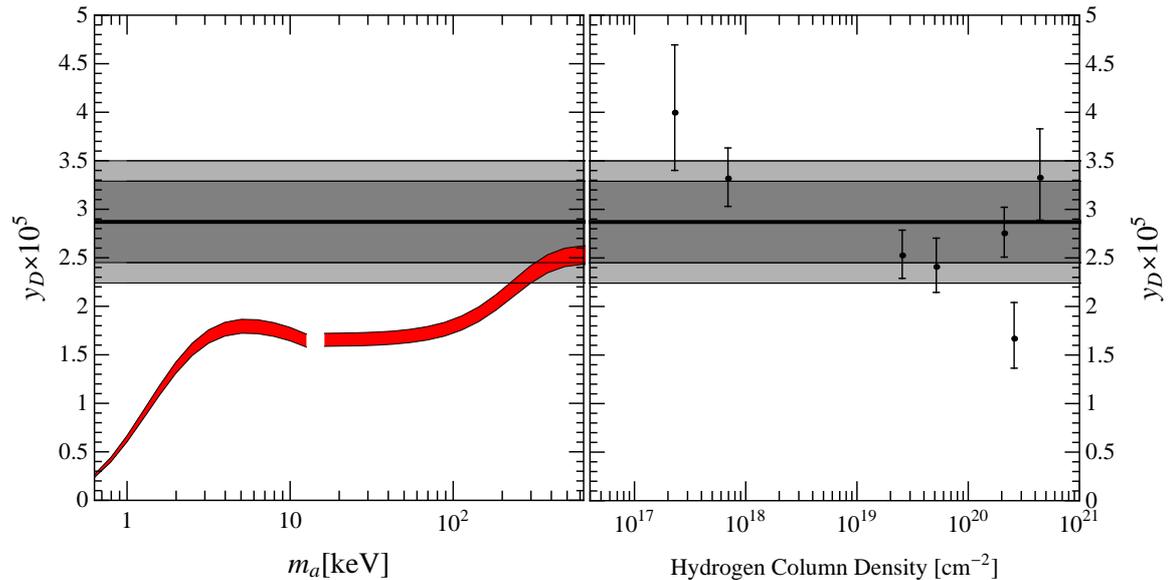}
   \caption{Deuterium yield $y_D$ as a function of $m_a$ for $\delta=1$ (Hadronic axions).
     The width of the red band represents the $1\sigma$ uncertainty
     of the CMB determination of $\eta$. To the right of the break,
     axions are treated as being in LTE, to the left they are assumed
     to be decoupled during BBN. The grey bands represent the 95\% and
     99\% range for the observed D abundance. It is derived from
     7 high redshift Ly-$\alpha$ clouds with individual results shown
     in the right panel as a function of the hydrogen column
     density~\cite{Pettini:2008mq}.}
   \label{fig4}
\end{figure}

In order to quantify our arguments we have modified the publicly
available BBN code \parte~\cite{Pisanti:2007hk} to include the
effects of axions, taking into account their impact on the Friedmann
equation and the modified densities of different radiation species.
Our results are shown in Fig.~\ref{fig4}.
For $m_a>20$~keV we treat axions as being in thermal equilibrium
throughout BBN. For $m_a\lesssim 10$~keV we use their abundance from our
numerical freeze-out calculation, assuming that they are decoupled
during BBN. We do not treat the intermediate case, leaving a gap in
the predicted deuterium yield as a function of $m_a$ that is shown as
a red band in Fig.~\ref{fig4}. To calculate this curve we have
adjusted the baryon abundance such that in the very end it matches
the CMB-implied value. Its $1\sigma$ range is represented by the
width of the red band.

We compare the predicted D yield with the measured value ${\rm
D/H}|_p =(2.82 \pm 0.21) \times 10^{-5}$~\cite{Nakamura:2010zzi},
estimated from 7 high-redshift, low-metallicity clouds absorbing the
light of background quasars~\cite{Pettini:2008mq}. The dark (light)
grey band in Fig.~\ref{fig4} is the experimental 95\% (99\%) C.L.\
region. We have included the independent determination of ${\rm
D/H}|_p$ from the 7 systems in the right panel of Fig.~\ref{fig4}
as function of the H column density. Note that there is a significant
scatter of the results whose origin still remains unclear and that
has been taken into account by artificially increasing the error (the
width of the grey band)~\cite{Pettini:2008mq}. Such a large 
scatter in the measurements of the D abundance may be a signal of some not yet 
understood processing of D inside these high-redshift clouds. Whether
this is the case or not, there are no known astration processes that increase the 
D concentration, so the primordial $y_{\rm D}$ should be \emph{larger} 
than this estimation. 
As we see in Fig.~\ref{fig4}, the presence of axions reduces $y_{\rm D}$, 
so our bounds appear to be conservative regarding this possible systematic.

The deuterium abundance is reduced below its 2$\sigma$ observation if
$m_a$ is below 300~keV. Therefore, BBN constrains axions to have
masses
\begin{equation}
m_a > 300\ {\rm keV}\,.
\end{equation}
Axions with mass above this limit have almost completely disappeared
from the thermal bath before they can affect BBN and the predictions
approach standard BBN.
Since this bound corresponds to axions that are always in thermal equilibrium,
even only via the Primakoff and inverse decay processes, it also applies to non-hadronic axions 
which would interact more strongly. 
Note finally that the helium yield is only mildly affected and
does not provide interesting bounds.

\section{Neutrino dilution}                           \label{sec:neff}

\subsection{Radiation density at CMB decoupling}

We have seen that thermally excited axions that eventually disappear
increase the abundance of photons at CMB decoupling relative to
baryons and neutrinos. Since all cosmic parameters are defined
relative to the observed CMB properties, this means that effectively
the neutrino abundance is reduced. The impact of this effect on the
CMB determination of the cosmic baryon abundance is minimal and thus
could be neglected in the previous section. However, precision
observables also measure the cosmic radiation density at decoupling,
and here the ``missing neutrinos'' should make a difference.

To extract cosmological information on the radiation density at CMB
decoupling we analyse the usual 8-parameter standard $\Lambda$CDM
model described in Ref.~\cite{Hannestad:2010yi}, extended in two
ways. We allow the effective number of neutrino degrees of freedom
to vary, assuming a flat prior on the interval $0< N_{\rm
eff}<3.0$. We recall that the radiation energy density is
traditionally expressed as
\begin{equation}\label{eq:rhorad}
\rho_{\rm rad}=
\left[1+\frac{7}{8}\,\left(\frac{4}{11}\right)^{4/3}N_{\rm eff}\right]
\,\rho_\gamma\,,
\end{equation}
where the photon radiation density is $\rho_\gamma=
(\pi^2/15)\,T^4$ and $N_{\rm eff}$ the effective number of
thermally excited neutrino degrees of freedom. The standard value is
$N_{\rm eff}=3.046$ instead of 3 because of residual neutrino heating
by $e^+e^-$ annihilation~\cite{Mangano:2005cc} but, given the current
experimental uncertainty, we neglect this tiny correction. In
addition, we allow for neutrino masses, assuming a common value
$m_\nu$ for all flavours and a flat prior on $0<\Omega_\nu/\Omega_{\rm
m}<1$. The other parameters and their priors are identical with those
provided in Ref.~\cite{Hannestad:2010yi}. Moreover, we use the same
set of cosmological data, i.e., the WMAP 7-year CMB measurements, the
7th data release of the Sloan Digital Sky Survey, and the Hubble
constant from Hubble Space Telescope observations.

\begin{figure}[t]
\hspace{25mm}
\includegraphics[width=10.cm,angle=0]{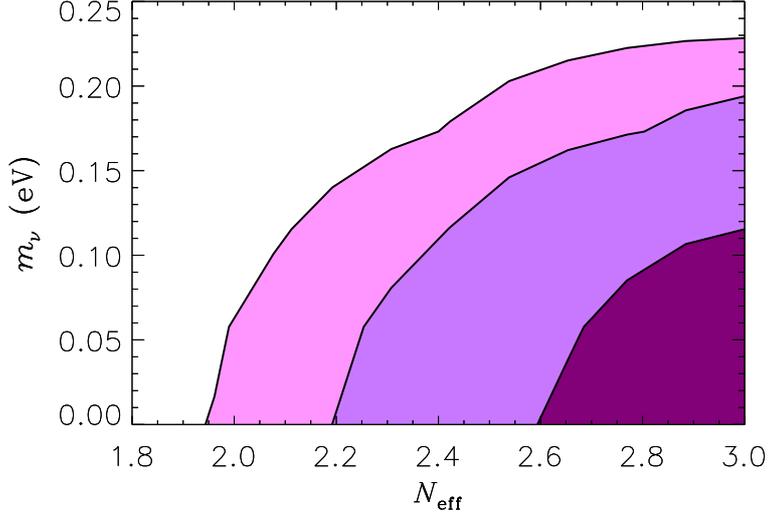}
\caption{2D marginal 68\%, 95\% and 99\% contours in the $m_\nu$--$N_{\rm eff}$
plane, where $m_\nu$ is the individual neutrino mass
(not the often-used sum over masses).
\label{fig5}}
\end{figure}

Marginalizing over all parameters but $m_\nu$ and $N_{\rm eff}$ we
find the 2D credible regions shown in Fig.~\ref{fig5}.
Marginalizing in addition over $m_\nu$ we find the limits
\begin{equation}\label{eq:neffconstraint}
N_{\rm eff}>
\begin{cases}
{2.70}&\hbox{at 68\% C.L.}\\
{2.39}&\hbox{at 95\% C.L.}\\
{2.11}&\hbox{at 99\% C.L.}
\end{cases}
\end{equation}
These limits are very restrictive because on present evidence
cosmology actually prefers extra radiation beyond $N_{\rm
eff}>3$
\cite{Hamann:2007pi,Hamann:2010pw,Komatsu:2010fb,GonzalezGarcia:2010un,Hamann:2010bk}.
Taking this possibility seriously, we would need to worry about {\em
two\/} novel ingredients: axions and radiation in some new form.

The PLANCK satellite, currently taking CMB data, is expected to boost
precision determinations of cosmological parameters. It will measure
the cosmic radiation content at CMB decoupling with a precision of
about $\Delta N_{\rm eff}=\pm0.26$ or better and thus will clearly
decide if there is extra radiation in the
universe~\cite{Hamann:2007sb}. If it finds convincing evidence for
extra radiation, much in cosmology will have to be reconsidered
besides our axion limits.

\subsection{Axion bounds}

\begin{figure}[tbp]
   \centering
   \includegraphics[width=7.3cm]{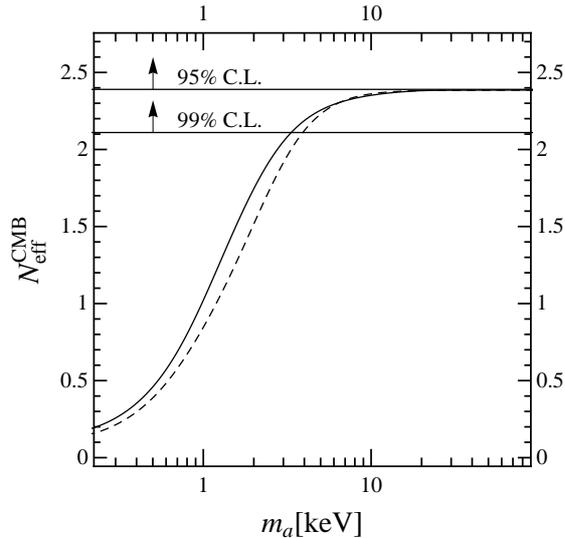}
   \caption{Radiation density during the CMB epoch ($\delta=1$).
   Solid and dashed lines are for hadronic and non-hadronic axions ($C_e=1/6$).
   The 95\% and 99\% C.L.\ lower limits from Eq.~(\ref{eq:neffconstraint})
   are shown as horizontal lines. }
   \label{fig6}
\end{figure}

The cosmic energy density in neutrinos is modified by the factor
$(T_\nu^{\rm ax}/T_\nu^{\rm std})^4$ between the axion and standard
cosmology. The results of Sec.~\ref{sec:freezeout} imply that this
ratio can be expressed in terms of the quantity $B_2$, the modified
$(T/T_\nu)^3$ value after axions have disappeared
\begin{equation}\label{eq:Neff}
N_{\rm eff}=3\(\frac{11}{4B_2}\)^{4/3}\,.
\end{equation}
We show the variation with $m_a$ for $\delta=1$ in
Fig.~\ref{fig6}. At sufficiently high $m_a$, the inverse
decay process keeps thermal equilibrium during the decay. Even if
axions decouple during $e^+e^-$ annihilation they will re-thermalize
with photons before decaying. In this case, we have an analytical
expression
\begin{equation}
N_{\rm eff} = 3\(\frac{11}{4}\frac{2(h_1+2)}{13}
\frac{2}{3}\)^{4/3}\(\frac{3}{2+h_1^{4/3}}\)\,.
\end{equation}
For $m_a>20$~keV we have $h_1\simeq 1$ and $N_{\rm eff}$ reaches
asymptotically the minimum neutrino dilution
$3\,(11/13)^{4/3}=2.401$.
At much larger masses, $m_a\sim$ MeV, axions would disappear in LTE
before neutrino decoupling, leaving no trace in cosmology. Therefore, at $m_a\sim$ MeV 
the value of $N_{\rm eff}$ shown in Fig.~\ref{fig6} 
reaches a plateau at the the standard value $N_{\rm eff}=3$.
Comparing with the cosmological limits of Eq.~(\ref{eq:neffconstraint}) we see that even the
minimum neutrino dilution is only barely allowed at 95\% C.L.\ and therefore disfavoured, but not credibly excluded.

For $m_a\alt 20$~keV, axions decay increasingly out of equilibrium,
creating entropy and reducing the final $N_{\rm eff}$ further.
Comparing the calculated $N_{\rm eff}$ in Fig.~\ref{fig6}
with the observational limits implies
\begin{equation}\label{eq:dilutionconstraint}
m_a>3~\rm keV~~\hbox{at~~99\% C.L.}
\end{equation}
The BBN limits reach to significantly larger masses, but the neutrino
dilution limits are still nicely complementary.

\section{CMB distortions}                      \label{sec:distortions}

The primordial plasma is optically thick before recombination, yet
the photons produced in axion decay can produce spectral distortions
in the CMB, measured by the FIRAS experiment to follow a black-body
spectrum with ${\cal O}(10^{-4})$ precision~\cite{Fixsen:1996nj}.

After $e^+ e^-$ annihilation, photon thermal equilibrium is
maintained by Compton scattering, double Compton scattering and
bremsstrahlung. Compton scattering is the fastest of these
processes,  but it can only provide kinetic equilibrium while
conserving the photon number density. Double Compton (DC) scattering
and bremsstrahlung (BR) are slower but change photon number and thus
achieve thermal equilibrium.
After a redshift of approximately $z \sim 3 \times 10^{6}$, DC and BR no longer occur
in equilibrium and any injection of photons leads to a non-zero pseudo degeneracy
parameter, $\mu$. The FIRAS data restrict its value to
\begin{equation}
\label{firas}
\mu <  9 \times 10^{-5}.
\end{equation}
Therefore, even a minute fraction of axion decays occurring after
this redshift leads to a measurable distortion.

Assuming that $\mu$ is always small, we can cast its evolution as~\cite{Hu:1992dc},
\begin{equation}
\label{muevolution}
\frac{d \mu}{d t} = \frac{d \mu_a}{dt}-\mu\(\frac{1}{t_{\rm DC}}+\frac{1}{t_{\rm BR}}\) ,
\end{equation}
where $t_{\rm DC}$ and $t_{\rm BR}$ are the $\mu$ relaxation time
scales due to the indicated processes (Appendix~\ref{app:freezeout})
and $d\mu_a/dt$ is the rate-of-change of $\mu$
generated by Compton scattering from photon injection or
disappearance resulting from axion decay or inverse decay. It is
related to the change in photon energy $d\rho_\gamma$ and photon
number $d n_\gamma$ by~\cite{Hu:1992dc}
\begin{equation}
\label{muinject}
d \mu_a=\frac{1}{2.14}\(\frac{3 d \rho_\gamma}{\rho_\gamma}-4\frac{d n_\gamma}{n_\gamma}\),
\end{equation}
assuming that $\mu$ is small. To obtain our bounds we have
implemented Eq.~\eqref{muevolution} in a numerical code describing
the evolution of axion decay (Appendix~\ref{app:freezeout}).

\begin{figure}[tbp]
   \includegraphics[width=7cm]{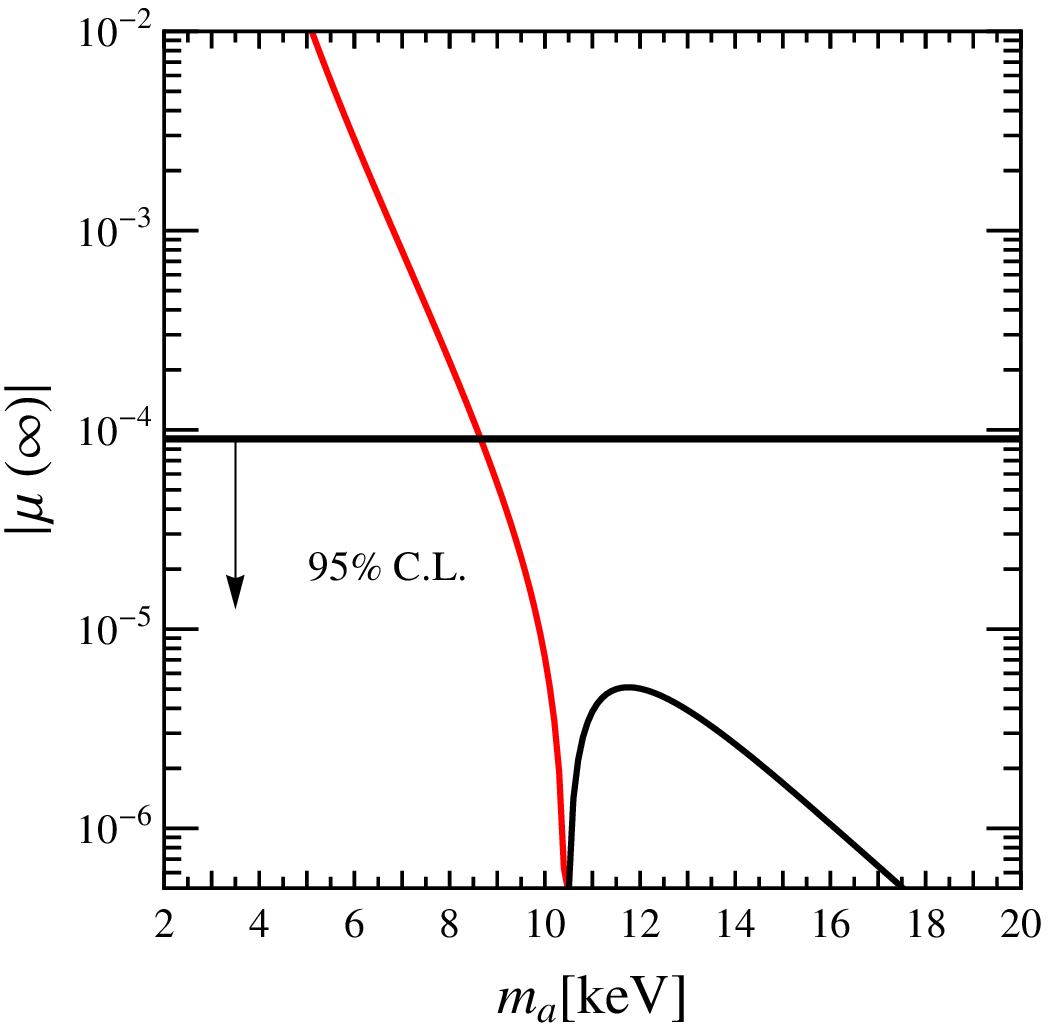}
   \includegraphics[width=7cm]{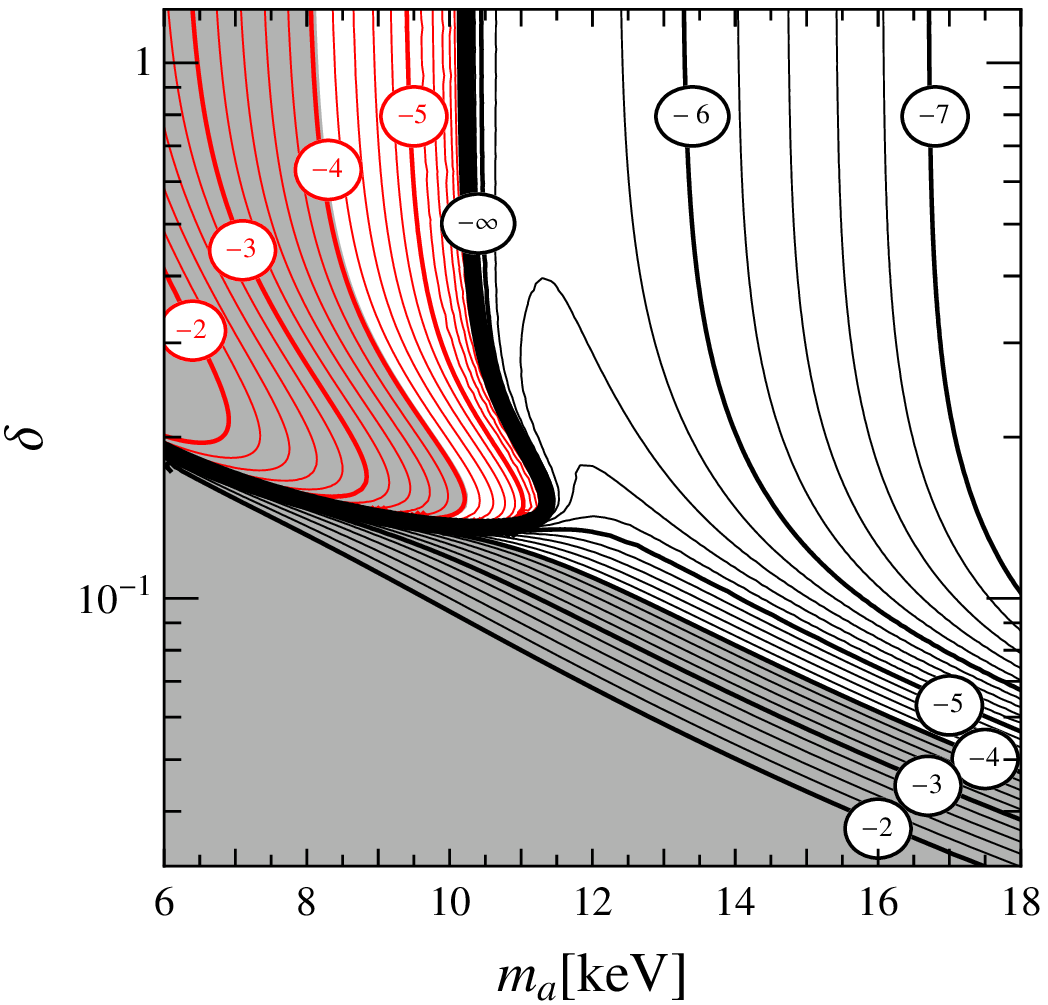}
   \caption{{\it Left:} Photon $\mu$ parameter after axion decay ($\delta=1$).
   The observational upper bound is indicated.
   {\it Right:}
   Contours of $\log_{10}|\mu|$ in the $\delta$--$m_a$--plane where
   black/red corresponds to positive/negative values of $\mu$
   (The thick black line corresponds to the boundary $\mu=0$).
   The shaded area is excluded.}
   \label{fig7}
\end{figure}

Our results are shown in Fig.~\ref{fig7}, where we plot the final
$\mu$ value as a function of $m_a$ and $\delta$ for hadronic axions. 
In the left panel we fixed $\delta=1$. In this case we find
\begin{equation}
m_a > 8.7~{\rm keV }~~\hbox{at~~95\%~C.L.}
\end{equation}
This bound is robust because $\mu$ is a steep function of $m_a$.
In particular it does not depend strongly on $h_1$ and therefore
is not expected to change significantly if we consider axions that couple to electrons, 
for which $h_1$ will be slightly larger.

The CMB distortion effect depends sensitively on the axion-photon
interaction strength for $\delta<1$ (right panel of
Fig.~\ref{fig7}). Generally the spectral distortions get larger
for smaller $\delta$ at a given $m_a$. For small $\delta$, the decay
happens later, when the photon distribution is less protected
against distortions.

For large $\delta$ the final $\mu$ changes sign from negative to
positive with increasing $m_a$ while for $\delta<0.1$ $\mu$ is always positive since axions decay
non-relativistically injecting more energy than photon number.
Of course, because of the sign change in $\mu$
there exist some fine-tuned cases where the final $\mu$ can be
accidentally zero.

Finally, we note that for photon injection occurring below $z \sim
10^5$ Compton scattering can no longer establish kinetic equilibrium,
and the spectral distortion no longer takes the shape of a
pseudo-degeneracy parameter. Rather, for $10^3 < z < 10^5$ it can be
represented by a Compton-$y$ parameter, and for $z < 10^3$ the decay
would show up as a bump in the infrared background. Most important
for our purpose here is that the bound on $\mu$ is weaker than on
other spectral distortions and therefore our bound is robust even for
$m_a$ well below 1~keV.

\section{Conclusions}                          \label{sec:conclusions}

Many cosmological particle-physics constraints were traditionally
based on the idea that the successful predictions for the
light-element abundances by BBN exclude additional radiation in the
universe. However, over the past few years it has become apparent
that both BBN and precision cosmology favour extra radiation on the
level of around one effective neutrino family. While this preference
is not yet strongly significant, it prevents one from deriving
restrictive limits on possible additional forms of radiation.

We have studied a scenario where BBN still provides useful
constraints on low-mass particles. Axions in the sub-MeV
mass range would produce photons late enough to dilute baryons after
BBN relative to photons. Therefore, the baryon abundance established
at CMB decoupling by cosmological precision data is smaller than at
BBN, leading to an unacceptably lowered deuterium abundance. We have
shown that this argument requires the mass of axions to
exceed around 300~keV. Additional arguments derive from the cosmic
radiation content at CMB decoupling (neutrinos would be diluted
relative to photons) and spectral CMB distortions.
These bounds depend sensibly on the two-photon coupling, which determines 
the axion decay, rather than on the couplings to leptons and hadrons.

In the case of hadronic axions, 
together with the hot-dark matter limit, we thus find that cosmology
alone excludes the $m_a$ range 0.7~eV--300~keV.  Our constraints are
complementary to well-known stellar evolution limits and laboratory 
bounds, but based on completely different reasoning.

\section*{Acknowledgements}

We acknowledge use of computing resources from the Danish Center for
Scientific Computing (DCSC). In Munich, partial support by the
Deutsche Forschungsgemeinschaft under the grant TR~27 ``Neutrinos
and beyond'' and the Cluster of Excellence ``Origin and Structure of
the Universe'' is acknowledged.

\appendix
\section{Axion out-of-equilibrium decay}         \label{app:freezeout}

\subsection{Boltzmann equation}

\begin{figure}[tbp]
\subfigure[]{\label{fig8a}\includegraphics[width=5cm]{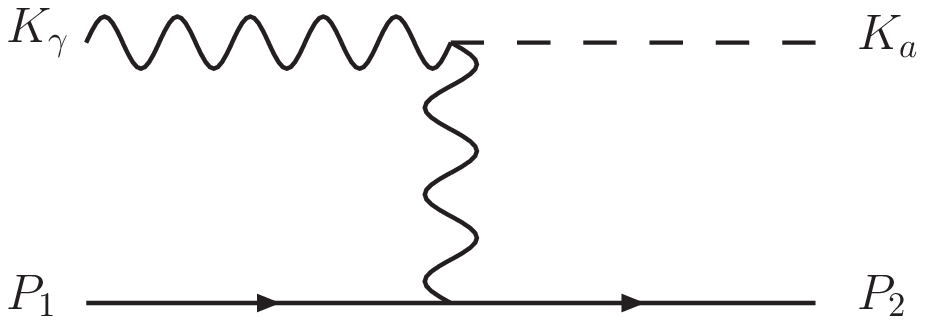}}
\hfill
\subfigure[]{\label{fig8b}\includegraphics[width=5cm]{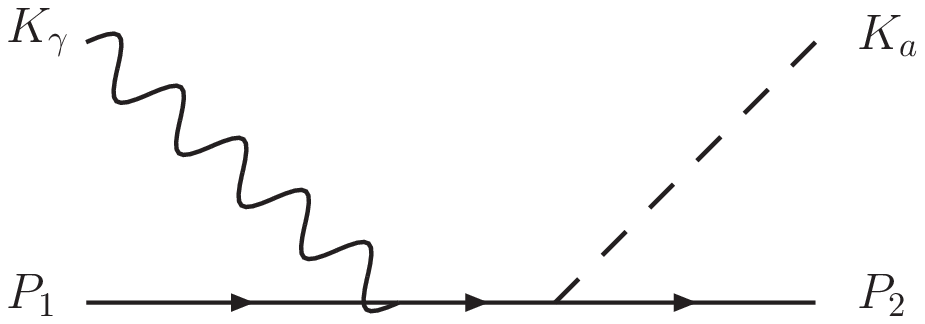}}
\hfill
\subfigure[]{\label{fig8c}\includegraphics[width=4cm]{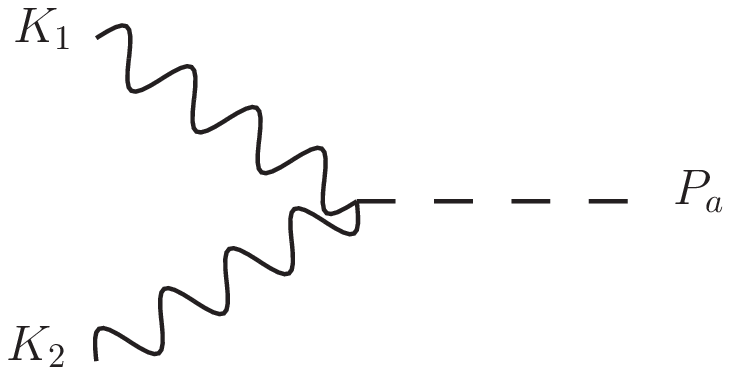}}
   \caption{Primakoff (a) and Compton (b) processes on electrons and/or positrons and inverse two-photon decay (c).
   In the Compton case, there is also an $u$-channel diagram contributing.}
   \label{fig8}
\end{figure}

We here compute the heating of the photon bath due to axion decay,
with concomitant neutrino and baryon dilution and induced spectral
$\mu$ distortion of the CMB. The relevant axion interaction processes
are Primakoff and Compton scattering on electrons and positrons and two-photon
decay and its inverse (Fig.~\ref{fig8}). The Boltzmann
equation for the axion phase space density $f_a(k_a,t)$ is
\begin{equation}\label{BOLfa}
\frac{\partial f_a}{\partial t}- k_a H \frac{\partial f_a}{\partial k_a}=
\left[C_e(k_a,f_a,T)+C_\gamma(k_a,f_a,T)\right]
(f_a^{\rm eq}-f_a)\,,
\end{equation}
where $H$ is the Hubble parameter. Here and in the following we
denote boson four vectors as $K=(\omega,{\bf k})$ and $k=|{\bf k}|$,
whereas for fermions we use $P=(E,{\bf p})$ and $p=|{\bf p}|$.
Specifically we use $\omega_a=(k_a^2+m_a^2)^{1/2}$. The collision
integrals are
\begin{eqnarray}
\nonumber
C_e &=&\frac{1}{2\omega_a}\int \frac{d^3{\bf p}_1}{(2\pi)^32E_1}
\frac{d^3{\bf p}_2}{(2\pi)^32E_2}
\frac{d^3{\bf k}_\gamma}{(2\pi)^32\omega_\gamma}\,
(2\pi)^4\delta^4(P_1+K_\gamma-P_2-K_a)\,|{\cal M}_e|^2
\label{caca1}\\
 &&\kern15em{}\times f_1(E_1)f_\gamma(\omega_\gamma)[1-f_2(E_2)]/f_a^{\rm eq}(\omega_a)\,,\\
 \nonumber
 C_\gamma &=&\frac{1}{2\omega_a}\int
 \frac{d^3{\bf k}_1}{(2\pi)^32\omega_1}\frac{d^3{\bf k}_2}{(2\pi)^32\omega_2}\,
 (2\pi)^4\delta^4(K_1+K_2-K_a)|{\cal M}_\gamma|^2
\label{caca2}\\
 &&\kern15em{}\times[1+f_1(\omega_1)+f_2(\omega_2)]\,,
\end{eqnarray}
where the particles are numbered as shown in Fig.~\ref{fig8}.
We assume thermal equilibrium between electrons and photons
(temperature $T$), neglecting possible chemical potentials (see
below). Relevant thermal effects are an effective photon mass,
preventing (inverse) decay at high $T$ and screening effects for the
Primakoff process. The spin- and polarization-summed squared matrix
elements are~\cite{Brodsky:1986mi}
\begin{eqnarray}
|{\cal M}_e|^2&=& |{\cal M}_P|^2+|{\cal M}_C|^2+2{\rm Re }\{{\cal M}^*_P{\cal M}_C\} \; , \\
|{\cal M}_P|^2&=& \frac{4 \pi \alpha g_{a\gamma}^2}{\(t-m_\gamma^2\)^2}
\Bigl\{-2 m_e^2 m_a^4-t \Bigl[m_a^4+2 \left(s-m_e^2\right)^2
-2 m_a^2 (s+m_e^2)\Bigr]
\nonumber\\
&&\kern6em{}-2t^2 \left(s-m_a^2\right)-t^3\Bigr\} \; , \label{mand} \\
|{\cal M}_C|^2&=& 16 \pi \alpha g_{ae}^2
\Bigl\{\frac{t^2-m_a^2(3 t-4m_e^4)}{(s-m_e^2)(m_e^2-u)}
-m_a^2\(\frac{s+m_e^2}{(s-m_e^2)^2}+\frac{u+m_e^2}{(u-m_e^2)^2}\) 
\Bigr\} , \\
2{\rm Re }\{{\cal M}^*_P{\cal M}_C\}&=& -16 \pi \alpha g_{ae} g_{a\gamma}
\frac{(m_a^2-t)^2}{(s-m_e^2)(m_\gamma^2-t)} \; ,
\\
|{\cal M}_\gamma|^2&=&\frac{1}{2}g_{a\gamma}^2m_a^2\left(m_a^2-4m_\gamma^2\right)\; ,
\end{eqnarray}
where we use the Mandelstam variables $s=(P_1+K_\gamma)^2$, 
$t=(K_\gamma-K_a)^2$ and $u=(K_a-P_1)^2$. 
Also we have used $g_{ae}= (C_e m_e)/f_a$. 
We keep the photon mass only in the propagator
to account for screening effects and on the external legs in
(inverse) decay to retain a threshold. For inverse decays we find
analytically
\begin{equation}\label{cgamma}
C_\gamma = \frac{m_a^2-4m_\gamma^2}{m_a^2}\frac{m_a}{\omega_a}
\[1+   \frac{2T}{p_a}\log
\frac{1-e^{-(\omega_a+p_a)/2T}}{1-e^{-(\omega_a-p_a)/2T}}\]\,
\Gamma_{a\to \gamma\gamma}\,.
\end{equation}

The Primakoff and Compton terms are much more complicated and have to be handled
numerically. In this case it is much simpler to compute axion
decoupling directly based on the Boltzmann equation for the axion
number density, cf.~Eq.~(\ref{BOLna}). Integrating Eq.~\eqref{BOLfa}
over the axion phase space and defining $Y_a=n_a/s$ (total entropy
$s$) we find
\begin{equation}\label{preYa}
s\frac{d Y_a}{dt}=\int \frac{d^3{\bf k}_a}{(2\pi)^3}
\(C_e+C_\gamma\)(f_a-f_a^{\rm eq}) \equiv (\Gamma_e+\Gamma_\gamma)
(n_a-n_a^{\rm eq})\,,
\end{equation}
where we have used comoving entropy conservation, $d(sR^3)/dt=0$, and
assumed that $f_a$ is proportional to $f_a^{\rm eq}$, a reasonable
approximation for computing the decoupling of a relativistic species.
The second equality defines the rates we have used in
Sec.~\ref{sec:freezeout}. Entropy conservation can also be used  to
obtain the temperature-time relation that allows us to use the former
as the independent variable
\begin{equation}\label{dt}
\frac{d}{dt}=-H\(\frac{d\log s}{d\log T^3}\)^{-1}
\frac{d}{d\log T}\,.
\end{equation}
which, together with~Eq.~\eqref{preYa}, leads to~Eq.~\eqref{BOLna}.

Our bounds do not depend on a precise determination of $h_1$, the axion
abundance after the $e^\pm$ annihilation epoch, so we have solved
Eq.~\eqref{BOLna} under two simplifying approximations. First, we
note that the axion contribution to $s$ is relatively small (the
contribution from photons and neutrinos are at least a factor 4.4
bigger) so we can neglect it in $s$ and in its derivative. Second, we
have assumed Maxwell-Boltzmann distributions for all particles to
find the average Primakoff/Compton rate, whereas decay and inverse decay are
tractable without this simplification\footnote{
We have checked that our results do not differ significantly with the full results 
derived in~\cite{Braaten:1991dd,Bolz:2000fu} for the relativistic limit $T\gg m_e,m_a$.}. It is then easy to obtain
numerically the evolution of $Y_a$ and the value of $h_1$ at
$T_{1}=m_e/10$.

\subsection{Axion decay}

At $T\sim m_e/10$ essentially all the $e^+e^-$ entropy has been
transferred to photons and axions. The $e^+e^-$ number density is
exponentially suppressed, so we can ignore Primakoff and Compton. To
compute numerically the photon heating when axions finally decay we
evolve the axion phase-space distribution. The initial condition is
set at $T=m_e/10$ by computing the axion temperature that gives
$n_a=h_1 n_a^{\rm eq}$ when assuming $f_a$ to be a thermal
distribution, albeit at a different temperature than photons. Note
that the Boltzmann equation~(\ref{BOLfa}) can be much simplified by
considering comoving momentum $\tilde{k}_a\equiv k_a(R/R_0)$, where
$R_0$ is a reference scale factor. Defining $g_a(\tilde{k}_a,t)\equiv
f(k_a,t)$, the l.h.s.\ of Eq.~(\ref{BOLfa}) simply is $dg_a/dt$ and
the Boltzmann equation describing the decay becomes
\begin{equation}\label{Boltzman1}
\frac{d g_a}{d t}=-\(g_a-g_a^{\rm eq}\)C_\gamma \; ,
\end{equation}
with $C_\gamma$ given by Eq.~(\ref{cgamma}).

We next compute the energy transfer to photons. The Boltzmann
equation for the photon phase space distribution
$f_\gamma(k_\gamma,t)$ can be written as
\begin{eqnarray}
\label{Boltzman2}
\frac{\partial f}{\partial t}-k_\gamma Hf &=&
\frac{1}{2\omega_\gamma}\int \frac{d^3{\bf k}_a}{(2\pi)^32\omega_a}
\frac{d^3k'_\gamma}{(2\pi)^32\omega_\gamma'}\,
(2\pi)^4\delta^4(K_a-K_\gamma-K_\gamma')\,|{\cal M}_\gamma|^2
\nonumber\\
 &&\kern10em{}\times\left[1+f_\gamma(\omega_\gamma)+f_\gamma(\omega_\gamma')\right]
 (f_a^{\rm eq}-f_a)
 \nonumber\\
 &&{}+ \hbox{interaction terms with relic baryons and electrons.}
\end{eqnarray}
The assumption that photons are always in kinetic equilibrium during
axion decay is equivalent to saying that the two sources on the
r.h.s.\ will readjust themselves such that $f_\gamma$ is always of
Bose-Einstein form. The evolution of the small photon chemical
potential is considered below and neglected in the axion decay
calculation.

Multiplying by the photon energy $\omega_\gamma$ and integrating over
phase space we obtain the equation for the evolution of the photon
energy density
\begin{equation}\label{energydensity}
\frac{\partial \rho_\gamma}{\partial t}-4 H \rho_\gamma =
\int \frac{d^3k_a}{(2\pi)^3} C_\gamma \left(g_a-g_a^{\rm eq}\right)\omega_\gamma\,.
\end{equation}
It states that the energy gain or loss of photons is due to axion
decays or inverse decays. We have neglected the terms involving
baryons and electrons. The evolution of the photon number density can
be obtained analogously, but one has to take into account that double
Compton and bremsstrahlung can yet modify the photon number (see
below).

To proceed further we define suitable time coordinates. As stated in
the main text, neutrino temperature $T_\nu$ provides a natural
reference to write the photon temperature in terms of $T_\nu$ through
the parameter $B$ that was defined in Eq.~(\ref{eq:AB}). Introducing
the variables $x=m_a/T_\nu$ and $y=P/T_\nu$ we have
\begin{equation}
\frac{d}{dt}=x H \frac{d}{dx}\; ,
\end{equation}
with $H$ defined in Eq.~(\ref{eq:Hubble}). The evolution of $g_a$ in
Eq.~(\ref{Boltzman1}) is in final form
\begin{equation}
\label{Boltzman3}
\frac{d g_a}{dx}=\frac{C_\gamma}{xH}(g_a-g_a^{\rm eq}) \ .
\end{equation}
Furthermore, the comoving photon energy density $\rho_\gamma
R^4=\pi^2 (T R)^4/15$ can be expressed in terms of $B^4$.
After some substitutions we find
\begin{equation}
\label{comovingT}
\frac{dB^4}{d x} = -  \frac{15}{\pi^2}
\int \frac{y^2 dy}{2\pi^2}
\frac{C_\gamma}{H}\frac{\sqrt{x^2+y^2}}{x} (g_a-g_a^{\rm eq})\,.
\end{equation}

So far we have ignored the effects of the photon chemical potential.
As explained in Sec.~\ref{sec:distortions} the evolution of a small
$\mu$ is given by Eq.~\eqref{muevolution}. The characteristic time
scales for double Compton scattering and bremsstrahlung to erase
$\mu$ are given in Ref.~\cite{Hu:1992dc}. Translated to our notation
they are
\begin{eqnarray}\label{TDCBR}
t_{\rm DC} &=&
2.1\times 10^{33}~{\rm s}\ \(1-Y_p/2\)^{-1}\(\Omega_b h^2\)^{-1}\(\frac{x T_{\gamma,{\rm today}}}{m_a}\)^{9/2}
\frac{B_{\rm today}^3}{B^{3/2}}\frac{1}{B_1^6} \; ,	\\ \nonumber
t_{\rm BR} &=&
3.4\times 10^{25}~{\rm s}\ \(1-Y_p/2\)^{-1}\(\Omega_b h^2\)^{-3/2}\(\frac{x T_{\gamma,{\rm today}}}{m_a}\)^{13/4}
\frac{B_{\rm today}^{9/2}}{B^{-5/4}}
\frac{1}{B_1^{27/4}}	\; , \\ \nonumber
\end{eqnarray}
with $Y_p$ the helium mass fraction and $\Omega_bh^2$ the baryon
density normalized to the critical density today (here $h$ stands for
the usual dimensionless Hubble constant). We use the values
$Y_p=0.23$, $T_{\gamma,{\rm today}}=2.725$ K and
$\Omega_bh^2=0.0223$. Using Eq.~\eqref{energydensity} and its
counterpart for photon number, only retaining the axion decay induced
$dn_\gamma$, the injection can be written as a time-dependent source
for  the pseudo degeneracy parameter
\begin{equation}
\label{chemicalmuevolution}
\frac{d \mu_a}{dx} = \frac{1}{2.14}\int \frac{dy y^2}{2\pi^2}\(g_a-g_a^{\rm eq}\)\frac{C_\gamma}{H}\frac{\sqrt{x^2+y^2}}{x}\(\frac{45}{\pi^2 B^4}-\frac{8\pi^2}{\zeta(3)B^3\sqrt{x^2+y^2}}\) .
\end{equation}

We have numerically solved the evolution equations for the axion
phase space distribution Eq.~\eqref{Boltzman3}, the photon
temperature Eq.~\eqref{comovingT}, and the pseudo degeneracy parameter 
Eq.~\eqref{muevolution}, together with the initial conditions
$\mu_0=0$  and for $B_1$ given by Eq.~\eqref{B1} as a function of
$h_1$. The latter was computed before, cf.\ Fig.~(\ref{fig2}). Note
that the expressions for $t_{\rm DC}$ and $t_{\rm BR}$,
cf.~Eq.~\eqref{TDCBR}, require the value of $B_{\rm today}$, so in
order to compute the chemical potential we first compute the
evolution of $B$ and determine $B_{\rm today}$ and then run again the
evolution to compute the final value of $\mu$.



\begin{thebibliography}{99}

\bibitem{Lesgourgues:2006nd} J.~Lesgourgues and S.~Pastor, ``Massive
    neutrinos and cosmology'', Phys.\ Rept.\  {\bf 429} (2006) 307 
    [astro-ph/0603494].

\bibitem{Hannestad:2006as} S.~Hannestad, H.~Tu and Y.~Y.~Y.~Wong,
    ``Measuring neutrino masses and dark energy with weak lensing
    tomography'', JCAP {\bf 0606} (2006) 025 [astro-ph/0603019].

\bibitem{Hannestad:2003ye} S.~Hannestad and G.~G.~Raffelt,
    ``Cosmological mass limits on neutrinos, axions, and other light
    particles'', JCAP {\bf 0404} (2004) 008 [hep-ph/0312154].

\bibitem{Hannestad:2005df} S.~Hannestad, A.~Mirizzi and
    G.~G.~Raffelt, ``New cosmological mass limit on thermal relic
    axions'', JCAP {\bf 0507} (2005) 002 [hep-ph/0504059].

\bibitem{Melchiorri:2007cd} A.~Melchiorri, O.~Mena and A.~Slosar,
    ``An improved cosmological bound on the thermal axion mass'',
    Phys.\ Rev.\  D {\bf 76} (2007) 041303 [arXiv:0705.2695].

\bibitem{Hannestad:2007dd} S.~Hannestad, A.~Mirizzi, G.~G.~Raffelt
    and Y.~Y.~Y.~Wong, ``Cosmological constraints on neutrino plus
    axion hot dark matter'', JCAP {\bf 0708} (2007) 015
    [arXiv:0706.4198].

\bibitem{Hannestad:2008js} S.~Hannestad, A.~Mirizzi, G.~G.~Raffelt
    and Y.~Y.~Y.~Wong, ``Cosmological constraints on neutrino plus
    axion hot dark matter: Update after WMAP-5'', JCAP {\bf 0804}
    (2008) 019 [arXiv:0803.1585]

\bibitem{Hannestad:2010yi}
  S.~Hannestad, A.~Mirizzi, G.~G.~Raffelt and Y.~Y.~Y.~Wong,
  ``Neutrino and axion hot dark matter bounds after WMAP-7'', 
  JCAP {\bf 1008} (2010) 001
  [arXiv:1004.0695].



\bibitem{Sikivie:2006ni} P.~Sikivie, ``Axion cosmology'', Lect.\
    Notes Phys.\  {\bf 741} (2008) 19 [astro-ph/0610440].

\bibitem{Raffelt:2006cw} G.~G.~Raffelt, ``Astrophysical axion
    bounds'', Lect.\ Notes Phys.\  {\bf 741} (2008) 51
    [hep-ph/0611350].

\bibitem{Arik:2008mq} E.~Arik {\it et al.}  (CAST Collaboration),
    ``Probing eV-scale axions with CAST'', JCAP {\bf 0902} (2009)
    008 [arXiv:0810.4482].

\bibitem{Inoue:2008zp} Y.~Inoue, Y.~Akimoto, R.~Ohta, T.~Mizumoto,
    A.~Yamamoto and M.~Minowa, ``Search for solar axions with mass
    around 1 eV using coherent conversion of axions into photons'',
    Phys.\ Lett.\  B {\bf 668} (2008) 93 [arXiv:0806.2230].

\bibitem{Masso:1995tw} E.~Mass\'o and R.~Toldr\`a, ``On a light
    spinless particle coupled to photons'', Phys.\ Rev.\  D {\bf
    52} (1995) 1755 [hep-ph/9503293].

\bibitem{Masso:1997ru} E.~Mass\'o and R.~Toldr\`a, ``New constraints
    on a light spinless particle coupled to photons'', Phys.\ Rev.\
    D {\bf 55} (1997) 7967 [hep-ph/9702275].

\bibitem{Aver:2010wq} E.~Aver, K.~A.~Olive and E.~D.~Skillman, ``A
    new approach to systematic uncertainties and self-consistency in
    helium abundance determinations'', JCAP {\bf 1005}, 003 (2010)
    [arXiv:1001.5218].

\bibitem{Izotov:2010ca} Y.~I.~Izotov and T.~X.~Thuan, ``The
    primordial abundance of 4He: evidence for non-standard big bang
    nucleosynthesis'', Astrophys.\ J.\  {\bf 710}, L67 (2010)
    [arXiv:1001.4440].

\bibitem{Hamann:2007pi}
 J.~Hamann, S.~Hannestad, G.~G.~Raffelt and Y.~Y.~Y.\ Wong,
 ``Observational bounds on the cosmic radiation density'',
 JCAP {\bf 0708} (2007) 021
 [arXiv:0705.0440].

\bibitem{Hamann:2010pw}
 J.~Hamann, S.~Hannestad, J.~Lesgourgues, C.~Rampf and Y.~Y.~Y.~Wong,
 ``Cosmological parameters from large scale structure---geometric versus shape
 information'',
  JCAP {\bf 1007}  (2010) 022
  [arXiv:1003.3999].

\bibitem{Komatsu:2010fb} E.~Komatsu {\it et al.}, ``Seven-Year
    Wilkinson Microwave Anisotropy Probe (WMAP) observations:
    Cosmological interpretation'', arXiv:1001.4538.

\bibitem{GonzalezGarcia:2010un}
  M.~C.~Gonzalez-Garcia, M.~Maltoni and J.~Salvado,
  ``Robust cosmological bounds on neutrinos and their combination with
  oscillation results'',
  JHEP {\bf 1008} (2010) 117 
  [arXiv:1006.3795].

\bibitem{Hamann:2010bk}
  J.~Hamann, S.~Hannestad, G.~G.~Raffelt, I.~Tamborra and Y.~Y.~Y.~Wong,
  ``Cosmology favoring extra radiation and sub-eV mass sterile
  neutrinos as an option'',
  Phys.\ Rev.\ Lett.\ {\bf 105} (2010) 181301
  [arXiv:1006.5276].


\bibitem{Bergsma:1985qz}
F.~Bergsma {\it et al.} [CHARM Collaboration],
``Search for axion like particle production in 400-GeV proton - copper interactions'',
Phys.\ Lett.\ B {\bf 157} (1985) 458.

\bibitem{Konaka:1986cb}
A.~Konaka {\it et al.},
``Search for neutral particles in electron beam dump experiment'',
Phys.\ Rev.\ Lett.\ {\bf 57 } (1986) 659.

\bibitem{Riordan:1987aw}
E.~M.~Riordan {\it et al.},
``A search for short lived axions in an electron beam dump experiment'',
Phys.\ Rev.\ Lett.\ {\bf 59} (1987) 755.

\bibitem{Bross:1989mp}
A.~Bross {\it et al.},
``A search for shortlived particles produced in an electron beam dump'',
Phys.\ Rev.\ Lett.\ {\bf 67 } (1991) 2942-2945.

\bibitem{Altmann:1995bw}
M.~Altmann {\it et al.},
``Search for the electron positron decay of axions and axion - like particles at a nuclear power reactor at Bugey'',
Z.\ Phys.\ {\bf C68 } (1995) 221-227.

\bibitem{Chang:2006ug}
H.~M.~Chang {\it et al.} [ TEXONO Collaboration ],
Phys.\ Rev.\ {\bf D75 } (2007) 052004.
[hep-ex/0609001].




\bibitem{Peccei:2006as}
 R.~D.~Peccei,
 ``The strong CP problem and axions'',
 Lect.\ Notes Phys.\  {\bf 741} (2008) 3
 [hep-ph/0607268].

\bibitem{Nakamura:2010zzi}
  K.~Nakamura (Particle Data Group),
  ``Review of particle physics'',
  J.\ Phys.\ G {\bf 37} (2010) 075021.

\bibitem{Kim:1979if}
  J.~E.~Kim,
  ``Weak interaction singlet and strong CP invariance'', 
  Phys.\ Rev.\ Lett.\  {\bf 43} (1979) 103.

\bibitem{Shifman:1979if}
  M.~A.~Shifman, A.~I.~Vainshtein and V.~I.~Zakharov,
  ``Can confinement ensure natural CP invariance of strong interactions?'', 
  Nucl.\ Phys.\  B {\bf 166}  (1980) 493.

\bibitem{Zhitnitsky:1980tq}
  A.~R.~Zhitnitsky,
  ``On possible suppression of the axion hadron interactions,'' (In Russian)
  Sov.\ J.\ Nucl.\ Phys.\  {\bf 31 } (1980)  260.

\bibitem{Dine:1981rt}
  M.~Dine, W.~Fischler and M.~Srednicki,
  ``A simple solution to the strong CP problem with a harmless axion'',
  Phys.\ Lett.\  {\bf B104 } (1981)  199.

\bibitem{Gelmini:1982zz}
  G.~B.~Gelmini, S.~Nussinov and T.~Yanagida,
  ``Does nature like Nambu-Goldstone bosons?'',
  Nucl.\ Phys.\  {\bf B219 } (1983)  31.  


\bibitem{Kolb:1990vq}
  E.~W.~Kolb and M.~S.~Turner,
  ``The early universe'',
  Front.\ Phys.\  {\bf 69} (1990) 1-547.

\bibitem{Pisanti:2007hk}
  O.~Pisanti, A.~Cirillo, S.~Esposito, F.~Iocco, G.~Mangano,
  G.~Miele and P.~D.~Serpico,
  ``PArthENoPE: Public Algorithm Evaluating the Nucleosynthesis of
  Primordial Elements'',
  Comput.\ Phys.\ Commun.\  {\bf 178}  (2008) 956
  [arXiv:0705.0290].

\bibitem{Pettini:2008mq}
  M.~Pettini, B.~J.~Zych, M.~T.~Murphy, A.~Lewis and C.~C.~Steidel,
  ``Deuterium Abundance in the Most Metal-Poor Damped Lyman alpha System:
  Converging on $\Omega_b$,"
  Mon.\ Not.\ R.\ Astron.\ Soc. {\bf 391} (2008)  1499Ð1510 
  [arXiv:0805.0594].

\bibitem{Mangano:2005cc}
G.~Mangano, G.~Miele, S.~Pastor, T.~Pinto, O.~Pisanti and P.~D.~Serpico,
``Relic neutrino decoupling including flavour oscillations'',
Nucl.\ Phys.\  B {\bf 729} (2005) 221
[hep-ph/0506164].

\bibitem{Hamann:2007sb}
 J.~Hamann, J.~Lesgourgues and G.~Mangano,
 ``Using BBN in cosmological parameter extraction from CMB: a forecast for
 Planck'',
 JCAP {\bf 0803} (2008) 004
 [arXiv:0712.2826].

\bibitem{Fixsen:1996nj}
 D.~J.~Fixsen, E.~S.~Cheng, J.~M.~Gales, J.~C.~Mather, R.~A.~Shafer and E.~L.~Wright,
 ``The cosmic microwave background spectrum from the full COBE/FIRAS data
 set'',
 Astrophys.\ J.\  {\bf 473} (1996)  576
 [astro-ph/9605054].

\bibitem{Hu:1992dc}
 W.~Hu and J.~Silk,
 ``Thermalization and spectral distortions of the cosmic background
 radiation'',
 Phys.\ Rev.\  D {\bf 48} (1993) 485. 

\bibitem{Brodsky:1986mi}
  S.~J.~Brodsky, E.~Mottola, I.~J.~Muzinich and M.~Soldate, 
  ``Laser induced axion photoproduction'',
  Phys.\ Rev.\ Lett.\  {\bf 56 } (1986)  1763.

\bibitem{Braaten:1991dd}
  E.~Braaten and T.~C.~Yuan,
  ``Calculation of screening in a hot plasma,''
  Phys.\ Rev.\ Lett.\  {\bf 66 } (1991)  2183-2186.

\bibitem{Bolz:2000fu}
  M.~Bolz, A.~Brandenburg and W.~B\"uchmuller,
  ``Thermal production of gravitinos,''
  Nucl.\ Phys.\  {\bf B606 } (2001)  518-544.

\end{thebibliography}
\end{document}